\documentclass[useAMS,usenatbib,usegraphicx]{mn2e}
\usepackage{amssymb}
\usepackage{lscape}

\title[New Wolf-Rayet and Weak Emission-Line CSPNe]{Newly Discovered Wolf-Rayet and Weak Emission-Line Central Stars of Planetary Nebulae}
\author[K. DePew et al.]{K. DePew,$^{1}$\thanks{E-mail: kyle.depew@mq.edu.au} Q.A. Parker,$^{1,2}$ B. Miszalski,$^{3}$ O. De Marco,$^{1}$ D.J. Frew,$^{1}$ A. Acker$^{4}$, \and A.V. Kovacevic$^{1}$ and R.G. Sharp$^{2}$ \\ 
$^{1}$Department of Physics and Astronomy, Macquarie University, Sydney, NSW 2109, Australia\\
$^{2}$Australian Astronomical Observatory, PO Box 296, Epping, NSW 1710, Australia\\
$^{3}$Centre for Astrophysics Research, University of Hertfordshire, College Lane, Hatfield AL10 9AB\\
$^{4}$Observatoire de Strasbourg, 11 Rue de l'Universit\'{e}, 67000 Strasbourg, France}

\begin{document}

\date{}

\pagerange{\pageref{firstpage}--\pageref{lastpage}} \pubyear{}

\maketitle

\label{firstpage}

\begin{abstract}
We present spectra of 32 previously unpublished confirmed and candidate Wolf-Rayet ([WR]) and weak emission-line (WELS) central stars of planetary nebulae (CSPNe).  Eighteen stars have been discovered in the Macquarie/AAO/Strasbourg H$\alpha$ (MASH) PN survey sample, and we have also uncovered 14 confirmed and candidate [WR]s and WELS among the CSPNe of previously known PNe.  Spectral classifications have been undertaken using both the Acker \& Neiner and Crowther, De Marco \& Barlow schemes.  Twenty-two members in this sample are identified as probable [WR]s; the remaining 10 appear to be WELS.  Observations undertaken as part of the MASH spectroscopic survey have now increased the number of known [WR]s by $\sim$30 per cent.  This will permit a better analysis of [WR] subclass distribution, metallicity effects, and evolutionary sequences in these uncommon objects. 
\end{abstract}


\section{Introduction}
Planetary nebulae (PNe) constitute a short-lived ($<$10$^{5}$ yr) phase in the life of intermediate mass ($\sim$1--8 M$_{\odot}$) stars \citep{i2}.  The majority of central stars of PNe (CSPNe) have hydrogen-rich atmospheres \citep{tgh}.  There does exist, however, another class of central stars with H-deficient atmospheres.  Some H-deficient CSPNe have strong emission lines of highly ionised carbon, oxygen and helium, and exhibit fast stellar winds and high mass-loss rates (up to 10$^{-6}$ M$_{\odot}$ per year) compared to other CSPNe.  These stars are known as Wolf-Rayet (denoted [WR]; \citealp{v1}) central stars (see Crowther 2008 for a summary of historical classification systems).  Quantitative classification schemes currently in use for [WR] subtypes are given by \citet{c1} and \citet{a2}.  

[WR] stars are often erroneously referred to as `Population II' WRs to differentiate them from their massive `Population I' counterparts.  (While most [WR]s are Population II objects, some are found within Type I PNe, which would indicate the CS to be Population I.)  Wind terminal velocities of up to 3000 km s$^{-1}$ have been measured in some cases, although late-type [WR]s have much smaller values, sometimes as low as $\sim$200 km s$^{-1}$ in [WC10] and [WC11]s \citep[e.g.][]{lhj96,dc99}.  There also exists a class of weak emission-line central stars (WELS; see \S\ref{ID}), which exhibit emission lines at the same wavelengths as Wolf-Rayet types, but at weaker intensities.  Unlike [WR]s, WELS do not necessarily have H-deficient atmospheres, though many do (Hajduk, Zijlstra \& Gesicki 2010).  Because of their similar spectra, our search for Wolf-Rayet type central stars in the MASH database has resulted in the concomitant discovery of a number of WELS, which are detailed along with the new Wolf-Rayet CSPNe.   


While there are nearly 3000 true, likely and possible Galactic PNe currently known \citep[e.g.][]{f1,jacoby10}, relatively few central stars have been spectroscopically classified.  For example, in the Strasbourg-ESO Catalogue of Galactic Planetary Nebulae \citep{a1}, out of 1143 PNe, only 224 central stars ($\sim$20 per cent) were assigned spectral types. More recently, \citet{weidmann11} have tabulated central star spectroscopic information for only 13 per cent of the enlarged Galactic PN sample. [WR] CSPNe are thought to represent between 5 and 7 per cent of all central stars \citep{tyl,t96,a2,frew10b}, but this may be a lower limit, based on a range of selection effects.  For example, when long-slit spectra are obtained for an object, the slit is often positioned over the brightest portion of the PN, missing the central star entirely.  For these reasons previously compiled samples may suffer from selection biases and the exact frequency of Wolf-Rayet type central stars could be higher than what has been suggested in the recent literature.  On the other hand, \citet{g4} find that [WR] PNe are probably the brightest PNe in H$\beta$ in a given sample, so in those samples where the faintest PNe are excluded, there exists the possibility of an overestimation of the frequency of [WR] stars.  Interestingly, the fraction of non-DA (hydrogen-deficient) white dwarfs in a local sample is approximately 40 per cent \citep{s4}.  This would suggest that the observed proportion of H-deficient central stars, including [WR]s and their progeny (e.g. PG 1159 stars; Werner \& Herwig 2006), the presumed progenitors of these non-DA white dwarfs, has been previously underestimated.

Observing CSPNe can be difficult.  Many are faint in the optical, as distance and interstellar reddening renders their observation problematic.  Because classification of [WR]s depends on measurement of sometimes faint optical lines, reddening will often hide the evidence of emission-line central stars unless deep, high S/N spectra are obtained.  

[WR] central stars are thought to result from a final late or very late helium shell thermal pulse after the star has left the AGB phase, which leads to the ingestion or ejection of the remaining thin hydrogen envelope \citep{h1,w5}.  This is effectively the same scenario used in the explanation of H-deficient ejecta inside otherwise normal PNe \citep[e.g. A 30 and A 78;][]{j2}, also called the `born-again' scenario \citep{i1}.  Arguments against the `born-again' scenario can be found in \citet{gt2000}, where the authors find a disparity between model surface brightness predictions and observed values.  The `born-again' scenario has also been questioned because the abundances of `born-again' ejecta in A 30 and A 58 have been found to be oxygen- instead of carbon-rich, with an overwhelming amount of neon, more akin to the abundances of neon in nova ejecta \citep{wesson03,wesson08,lau10}.  Alternative theories have been proposed for the evolutionary sequence of [WR] stars, among them a binary interaction \citep{d2,g3,d4}.  
\citet{hzg} recently found that the first close binary [WR] CSPN and suggested that the close binary fraction in the [WR] class may be much lower than for their H-rich counterparts. 

Our incomplete understanding of [WR] evolution is due in part to the dearth of objects in this class which may be examined.  
  Hence, any new additions to the list of known [WR]s are important in constructing a rigorous sample for statistical analysis.  \citet{g2} and \citet{a2} show that very early-type ([WO1]-[WO4]) and late-type ([WC8]-[WC11]) stars dominate [WR] CSPNe in the Galaxy, in contrast to the Magellanic Clouds, which seem to possess primarily intermediate types between [WC4] and [WC8] \citep{p3}.  Recent work has revealed that the Galactic bulge [WR]s seem to be heavily weighted towards late spectral subtypes \citep{g1}.  In this respect, the bulge [WR] subtype distribution is similar to massive WC stars, which seem to increase in number as the subtype number increases \citep{crowther}. 

With a larger sample, it will be possible to put more precise limits on the frequency of [WR]s exhibiting specific subclasses and explore the differences in type distribution between environments of differing metallicity and age.  In addition, a greater variation of parameters such as metallicity, temperature, terminal velocity and chemical abundances can be useful in theoretical work which attempts to place constraints on the conditions necessary for H-deficient stellar evolution.  In the future, separate samples for the Galactic disk (intermediate age, metal-rich) and bulge (old, metal-rich), as well as the Magellanic Clouds (intermediate age, metal poor) 
may enable us to more precisely lay out the age and metallicity constraints required for the specific subclasses to be produced.
      


\subsection{[WR]s in the MASH Sample}

The number of known Galactic [WR] CSPNe has steadily increased from the late 1960s, when such nuclei were first discovered and classified \citep{s1}. By the mid-1990s, the count had reached 56  \citep{j1}. Discoveries over the past ten years, including our work, have nearly doubled this number to approximately 105.  \citet{a2} summarised the state of the field early in the decade, listing 86 objects designated to be a [WR] or a [WR] candidate.  

The Macquarie/AAO/Strasbourg H$\alpha$ (MASH) surveys of PNe \citep{p1,m1} presented $\sim$900 and $\sim$300 new PNe respectively.\footnote{Note the MASH-II sample has recently increased to $\sim$360 PN candidates.}  In all, the MASH survey increased the number of known Galactic PNe by nearly 80 per cent.  Out of the MASH catalogues, 24 [WR] stars or [WR] candidates have been identified. Only seven of these [WR]s have been previously published in detail, in three previous papers.  Morgan et al. (2001; Paper I), detailed the first two discoveries, while another four [WR] central stars were presented by \citeauthor{p2} (2003; Paper II).  A paper on the purported [WN] PM~5 (see \S\ref{ID}) quickly followed \citep[][Paper III]{m3}.  This paper represents the fourth in our series concerning new [WR] discoveries, where we describe our additional discoveries in detail and present their spectra.  Many of them were listed in brief notes in the online MASH catalogues\footnote{See http://vizier.cfa.harvard.edu/viz-bin/VizieR?-source=V/127} accompanying Parker et al. (2006) and Miszalski et al. (2008a).  Note that the preliminary discovery spectrum for MPA1611-4356 was shown  by Miszalski et al. (2008a) with others noted in the catalogue and referred to generally (but not explicitly) in Miszalski, Acker \& Parker (2008b).   Four of the objects noted here have been independently identified as [WR]s by \citet{s5} and \citet{g4}, but our observations and identification predate these publications.  We note the respective references in the specific subsections devoted to each object.  


\section{Spectroscopic Observations}
\label{SpecObs}

Spectroscopic observations of most objects were carried out at the Siding Spring Observatory on the Anglo-Australian Telescope, the 2.3m telescope, and the UK Schmidt telescope (UKST).  Other data were taken at the South African Astronomical Observatory (SAAO) using the 1.9m telescope and at the European Southern Observatory's Very Large Telescope during the ESO visitor mode programme 0.79.D-0764(A).  All spectra were bias and flat-field subtracted and dereddened, and in some cases  strong nebular lines from the surrounding PN  were also subtracted.  Those objects with signal-to-noise ratios (SNR) of 10 or greater were rectified.  The instrument and observation details concerning the discovery spectra are listed in Table \ref{ObsInfo}.  For long slit observations, the slit width was set at 2''.  Fibres on AAT 2dF observations are 2.1'' wide.  Observations on 6dF on the UKST used 6.7'' fibres. 

As discussed in Section 1.1, the majority of the [WR] central stars presented here were identified in the course of spectroscopically confirming MASH PN candidates when the fibre or slit fell across the CSPN.  In addition, 10 previously known PNe were serendipitously discovered to possess [WR] CSPN in the course of general MASH follow-up observations with the AA$\Omega$ spectrograph \citep{aao} on the Anglo-Australian Telescope and the 6dF spectrograph \citep{6df} on the UKST.  The object of observing missions with these wide-field multi-object spectrographs was to confirm as many MASH PNe as efficiently as possible.  These multi-object spectrograph (MOS) instruments possess 400 and 150 fibres over a two degree and a six degree field respectively.  This enables us to place individual fibres not just on the primary MASH candidate objects but also on all other known PNe that are present in the field.  Subsequently, during the long exposure times required to obtain sufficient SNR in observations of the faint MASH PNe, the faint [WR] features of some central stars of previously known PNe also emerged.  For a description of the AA$\Omega$ observations, see \citet{misz09}.

Spectra of most of the 32 [WR]/WELS central stars presented here are to be found in Figs. \ref{Spec1}, \ref{Spec2} and \ref{Spec3}.  

\begin{landscape}
\begin{table}
\begin{minipage}{170mm}
\caption{Observational details of the new CSPNe discovery spectra.}
\begin{tabular}{l l c c c c c c c c c}
\hline
PN G & RA & Dec & Name & Observatory$^{1}$/ & Instrument & Exposure & Grating & R & SNR$^{2}$ & Date Observed \\ [0.5ex]
& & & & Telescope & & Time (s) & & & & \\
\hline
\hline
001.2-05.6 & 18:11:02.7 & -30:42:11 & PHR1811-3042 & SSO/AAT & AA$\Omega$-SPIRAL & 1200 & 385R/580V & 1300 & 5 & 28 Jun 2006 \\
016.8-01.7 & 18:27:50.8 & -15:04:23 & BMP1827-1504 & ESO/VLT-U2 & FLAMES & 2400 & LR2-3/LR5-7 & 8900-12000 & -- & 13 Jun 2007 \\
037.7-06.0 & 19:21:44.5 & +01:32:41 & MPA1921+0132 & SSO/2.3m & DBS & 300 & 300B/1200R & 1200/7000 & -- & 13 May 2007 \\
216.0+07.4 & 07:23:48.4 & +00:36:32 & PHR0723+0036 & SSO/2.3m & DBS & 1200 & 300B & 1200 & 20 & 15 Nov 2007 \\
222.8-04.2 & 06:54:13.4 & -10:45:38 & PHR0654-1045 & SAAO/1.9m & CCD SPEC & 600 & 300B (G7) & 900 & 15 & 28 Jan 2003 \\
284.2-05.3 & 10:01:18.8 & -61:52:04 & PHR1001-6152 & SAAO/1.9m & CCD SPEC & 1200 & 300B (G7) & 900 & 20 & 01 Feb 2003 \\
297.0+06.5 & 12:09:29.1 & -55:53:33 & BMP1209-5553 & SSO/2.3m & DBS & 900 & 300B & 1200 & -- & 18 Feb 2007 \\
302.0-01.6 & 12:43:19.4 & -64:28:01 & MPA1243-6428 & SSO/2.3m & DBS & 600 & 300B & 1200 & -- & 19 Feb 2007 \\
309.8-01.6 & 13:54:22.4 & -63:37:18 & MPA1354-6337 & SSO/2.3m & DBS & 600 & 300B & 1200 & -- & 22 Feb 2007 \\
313.4+06.2 & 14:05:32.2 & -55:07:44 & MPA1405-5507 & SSO/2.3m & DBS & 600 & 300B & 1200 & 15 & 22 Feb 2007 \\
313.9+02.8 & 14:16:37.7 & -58:09:32 & PHR1416-5809 & SAAO/1.9m & CCD SPEC & 1200 & 300B (G7)  & 900 & 5 & 22 Jul 2002\\
331.8-02.3 & 16:24:02.9 & -52:50:05 & MPA1624-5250 & SSO/2.3m & DBS & 3600 & 300B & 1200 & 5 & 13 May 2007 \\
336.5+05.5 & 16:11:12.9 & -43:56:22 & MPA1611-4356 & SSO/2.3m & DBS & 1200 & 300B & 1200 & 5 & 23 Feb 2007 \\
348.4+04.9 & 16:55:22.0 & -35:35:24 & MPA1655-3535 & SSO/2.3m & DBS & 300 & 300B & 1200 & 5 & 28 Feb 2007 \\
355.9-04.4 & 17:53:39.8 & -34:43:40 & PHR1753-3443 & SSO/UK Schmidt & 6dF & 3 x 600 & 425R & 1000 & 10 & 21 Aug 2003 \\
356.0-04.2 & 17:53:04.9 & -34:28:39 & PHR1753-3428 & SSO/UK Schmidt & 6dF & 3 x 600 & 425R & 1000 & 5 & 21 Aug 2003 \\
356.8-03.6 & 17:52:29.2 & -33:30:04 & PHR1752-3330 & SSO/AAT & AA$\Omega$-2dF & 1800 & 385R/580V & 1300 & 20 & 8 Aug 2008 \\
359.8+03.5 & 17:31:47.6 & -27:09:18 & PHR1731-2709 & SSO/AAT & AA$\Omega$-2dF & 1800 & 385R/580V & 1300 & -- & 30 May 2008 \\
\hline
000.7-02.7 & 17:58:09.6 & -29:44:20 & M 2-21 & SSO/AAT & AA$\Omega$-2dF & 1800 & 385R/580V & 1300 & -- & 30 May 2008 \\
003.9-02.3 & 18:03:39.3 & -26:43:34 & M 1-35 & SSO/AAT & AA$\Omega$-2dF & 1800 & 384R/580V & 1300 & 10 & 29 May 2008\\
008.1-04.7 & 18:22:01.1 & -24:10:40 & M 2-39 & SSO/UK Schmidt & 6dF & 1200 & 385R/580V & 1000 & 25 & 11 Aug 2004 \\
008.2-04.8 & 18:22:32.0 & -24:09:28 & M 2-42 & SSO/UK Schmidt & 6dF & 1200 & 385R/580V & 1000 & 10 & 20 Aug 2004 \\
029.0+00.4 & 18:42:46.9 & -03:13:17 & Abell 48 & SSO/2.3m & DBS & 300 & 1200B/1200R & 5000/7000 & -- & 11 May 2008 \\
270.1-02.9 & 08:59:02.9 & -50:23:40 & Wr 17-23 & SSO/2.3m & DBS & 1200 & 300B & 1200 & 5 & 17 Nov 2007 \\
322.4-00.1 & 15:23:42.9 & -57:09:25 & Pe 2-8 & ESO/VLT-U2 & FLAMES & 1800 & LR3/LR5-7 & 8900-12000 & 5 & 11 Jun 2007 \\
355.4-04.0 & 17:51:12.1 & -34:55:22 & Pe 1-10 & SSO/AAT & AA$\Omega$-2dF & 1200 & 385R & 1300 & 5 & 27 Mar 2007 \\
356.1+02.7 & 17:25:19.4 & -30:40:42 & Th 3-13 & SSO/AAT & AA$\Omega$-2dF & 1800 & 385R & 1300 & -- & 30 May 2008 \\
356.9+04.4 & 17:21:04.4 & -29:02:59 & M 3-38 & ESO/VLT-U2 & FLAMES & 1200 & LR2-3/LR5-7 & 8900-12000 & -- & 10 Jun 2007 \\
357.4-03.2 & 17:52:34.4 & -32:45:52 & M 2-16 & SSO/AAT & AA$\Omega$-2dF & 2 x (1800/1500) & 385R/580V & 1300 & 5 & 08 Aug 2008 \\
358.0-04.6 & 17:59:55.0 & -32:59:12 & Sa 3-107 & SSO/AAT & AA$\Omega$-2dF & 1800 & 385R/580V & 1300 & 10 & 08 Aug 2008 \\
358.9+03.3 & 17:30:02.5 & -27:59:18 & H 1-19 & SSO/AAT & AA$\Omega$-2dF & 1800 & 385R/580V & 1300 & 5 & 30 May 2008 \\
359.9-04.5 & 18:03:52.6 & -31:17:46 & M 2-27 & SSO/AAT & AA$\Omega$-2dF & 1800 & 385R/580V & 1300 & 5 & 30 May 2008 \\
\hline
\hline
\label{ObsInfo}
\end{tabular}
\begin{flushleft}
$^{1}$ESO=European Southern Observatory-La Silla, SAAO=South African Astronomical Observatory, SSO=Siding Spring Observatory.\\
$^{2}$SNR was measured in the continuum at $\sim$5900\AA.  When no SNR is indicated this means that the continuum was not detected.\\
\end{flushleft}
\end{minipage}
\end{table}
\end{landscape}

\section{Classification Schemes}
\label{ID}

Wolf-Rayet central stars of PNe are divided into [WC] (carbon-sequence), [WO] (oxygen-sequence) and now a putative intermediate [WN/WC] class \citep[e.g. PB8;][]{t1}.  LMC-N66 \citep{N66} and PM~5 \citep{m3} have also been classed as [WN] types, which differ from the [WN/WC] class only in carbon depletion.  However, PM~5 may also be a massive WR ring nebula \citep{f1}.   It should be noted that [WO] and [WC] types are named according to their most prominent lines and not because of any significant difference in chemical abundances \citep{c1}.  The existence of the [WN/WC] and [WN] classes is still controversial.

The [WC] and [WO] subtype classifications are assigned according to the scheme of \citeauthor{c1} (1998; hereafter CDB98) and the system devised by \citeauthor{a2} (2003; hereafter AN03). CDB98 refined the WO and WC schemes used for massive stars, and proposed a unified classification system for both massive WRs and [WR]s.  AN03 extended the CDB98 criteria to a larger spectral range.  AN03 constructed a grid of selected line-intensities ordered by decreasing ionisation potential going from 871 to 24 eV. In this grid, the stars appear to belong clearly to prominent O (hot [WO1-4] types) or C (cooler [WC4-11] types) line-sequences, in agreement with the classification of massive WR stars applied to CSPN by Crowther et al. 1998 (CMB98). AN03 used 20 selected line ratios and the FWHM of CIV and C III lines as classification diagnostics. 

Both classification schemes emphasize the identification of the C IV doublet at 5801 and 5812\AA.  This doublet should be a broad feature (EW$>$10\AA) arising from the rapid expansion of the hot stellar wind.  When the doublet has a small equivalent width and the 4650\AA\ N III/C III-IV complex is present in a spectrum, the central star may be classified as a weak emission-line star (WELS) \citep{tyl,m4}.  These objects generally have narrower lines than [WR]s and are highly ionised, sometimes to such a degree that the C IV $\lambda\lambda$5801,12 doublet is present but the C III $\lambda$5696 line is not \citep{tyl}.  Because the C IV lines are so weak, they are often resolved into the separate components at 5801 and 5812\AA.  There also exists a 'very late' (VL) classification for those [WC] stars that would meet the [WC11] criteria and also exhibit a C II $\lambda$7236 line \citep{g4}.  These objects generally reside in low-ionisation PNe.   

For more precise [WR] classification, various C and O lines must be measured and their dereddened equivalent width or flux ratios determined.  However, the majority of these features are relatively weak, often leaving the exact subclass indeterminate in the absence of reasonably high S/N spectra.  Deeper observations will be required to identify the exact spectral type for the majority of our newly discovered [WR] central stars.

In this paper each object's classifications according to the CDB98 and AN03 schemes are listed in Table \ref{New[WR]s}.   Classification in both schemes generally agrees, though one scheme is sometimes more precise than the other.  The [WC] classification in both schemes depends primarily on the presence of the C IV $\lambda\lambda$5801,12 doublet and the C III $\lambda$5696 line.  Other lines, namely O VI $\lambda\lambda$3811,34 and O VII $\lambda$5670, are necessary for classification as a [WO] star.  See \citet{c1} and \citet{a2} for example spectra for each subclass.

\section{Individual Objects}

\begin{figure*}
\setlength{\intextsep}{1.0cm}
\begin{minipage}[t]{20cm}
\begin{center}
\includegraphics[clip=true,scale=0.85]{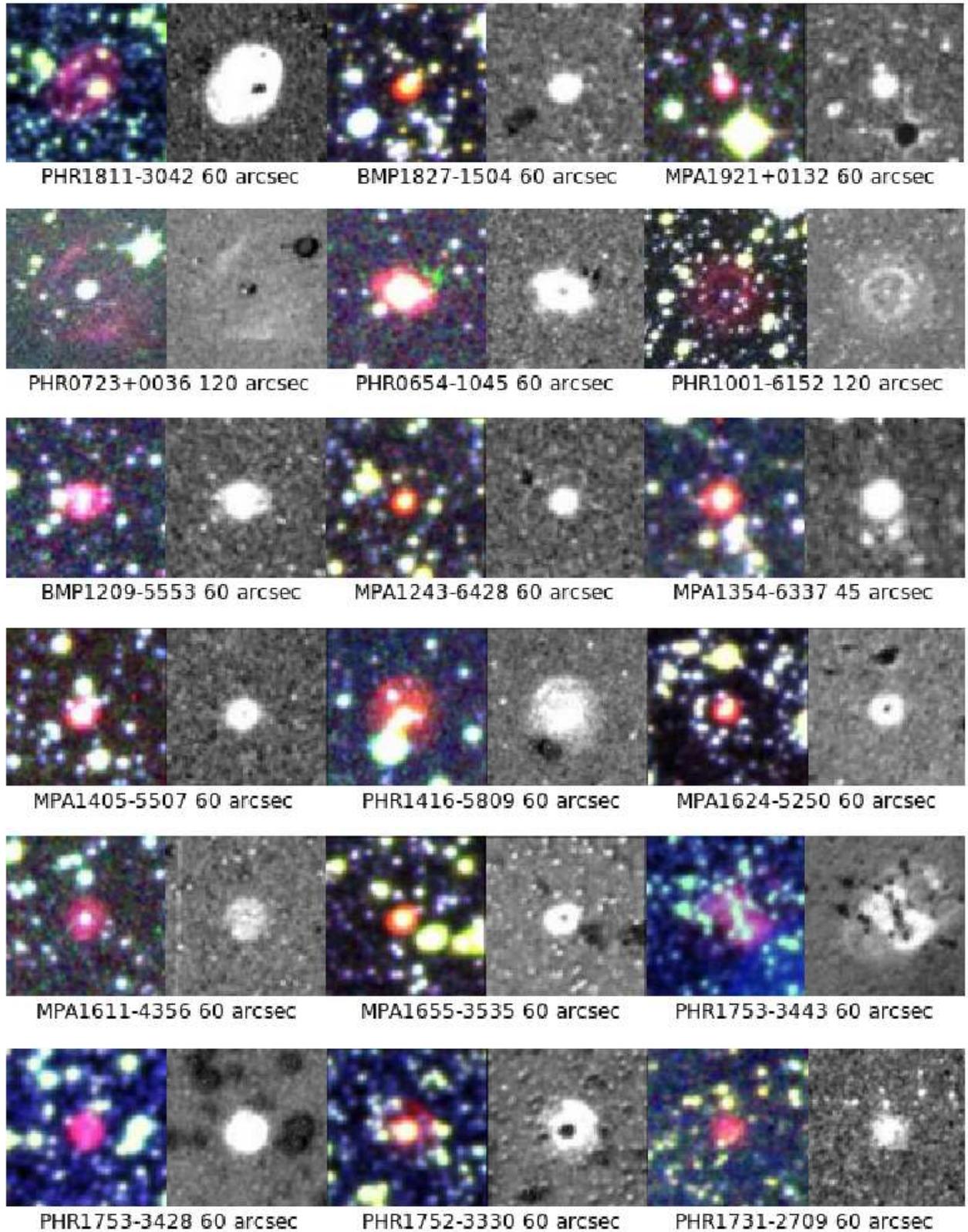} 
\end{center}
\end{minipage}
\caption{\setlength{\baselineskip}{20pt}A montage of the new MASH [WR] and WELS PNe, ordered according to Galactic longitude.  Each H$\alpha$/SR/B$_{J}$ composite colour image is accompanied by the H$\alpha$/short-red quotient image to its right.  The H$\alpha$/SR images are from the SuperCOSMOS H$\alpha$ Survey \citep{park05} and the B$_{J}$ images from \citet{ham}.  The lengths of the image sides in arcseconds are presented alongside the name of each object.  North is to the top and east is to the left for all images.}
\begin{flushleft}
\end{flushleft}
\label{PNMontage}
\end{figure*}

\begin{figure*}
\setlength{\intextsep}{1.0cm}
\begin{minipage}[t]{20cm}
\begin{center}
\includegraphics[clip=true,scale=0.85]{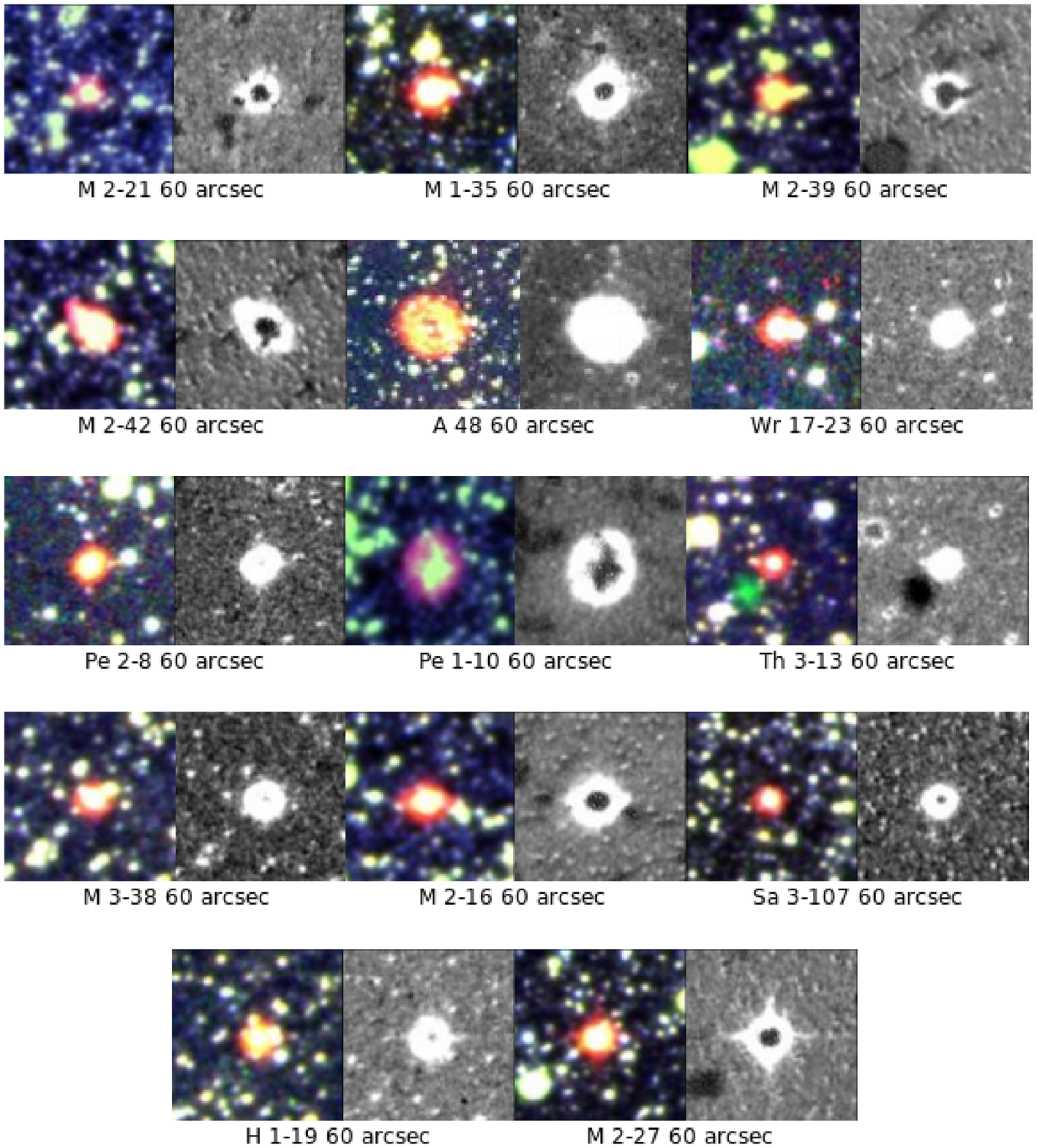} 
\end{center}
\end{minipage}
\caption{\setlength{\baselineskip}{20pt}A montage of the non-MASH PNe found to contain a true or candidate [WR] or WELS central star.  As in Fig. \ref{PNMontage}, each H$\alpha$/SR/B$_{J}$ composite colour image \citep{park05,ham} is accompanied by the H$\alpha$/short-red quotient image to its right.  Again, the lengths of the image sides in arcseconds are presented alongside the name of each object.  North is to the top and east is to the left for all images.}
\label{OldPNMontage}
\end{figure*}

Below we provide details of the spectra of the new [WR] CSPNe, separated into those discovered in MASH PNe and in other previously known PNe.  See Table \ref{ObsInfo} for observation details and Table \ref{New[WR]s} for [WR] classification, morphology and photometric data.  Images of the MASH [WR]s and [WR] candidates taken from the AAO/UKST H$\alpha$ Survey and the SuperCOSMOS Sky Survey \citep{park05,ham} can be seen in Fig. \ref{PNMontage}.  Those of the non-MASH [WR] objects can be seen in Fig. \ref{OldPNMontage}.

\subsection{New [WR] Stars}

In this section we briefly detail the characteristics of the [WR] central stars of our PN sample, ordered by increasing Galactic longitude.

\subsubsection{PHR1811-3042} 

The [WR] nature of the CSPN of PHR1811-3042 was discovered during general MASH candidate follow-up service time on the AA$\Omega$ system at the Anglo-Australian Telescope.  C IV $\lambda\lambda$4658,7726; O IV $\lambda$7032,53; O VI $\lambda\lambda$3811,5290 and He II $\lambda\lambda$4686,5411 are visible in the spectrum presented by \citet{s3}.  The C III $\lambda$5696 line is not evident.  The prominent oxygen spectrum and high ionisation stage of all lines suggests that this star is a [WO].  However,  the C IV doublet at $\lambda\lambda$5801,12 is not visible, likely due to the absence of a continuum in that spectral region.  The impossibility of measuring the doublet means that we cannot classify this star further.

\subsubsection{BMP1827-1504}

BMP1827-1504 was mentioned in \citet{c2} as a candidate PN and confirmed in MASH-II.  We have only observed the spectral ranges 3960-5007 and 5700-8000\AA\ (Figs. \ref{Spec1} to \ref{Spec3}).  The C IV $\lambda\lambda$5801,12 doublet appears to be the only stellar feature in the spectrum; it is weak but broad.  The continuum of the spectrum is undetected, which means that the absence of a line does not mean that the line is not present. In addition, C III $\lambda$5696 resides in an unobserved part of the spectrum.  We can therefore not carry out the usual comparison between the nearby C III $\lambda$5696 and C IV $\lambda\lambda$5801,12, which even in the absence of a continuum can give an indication of the ionisation state of the star.  Only a general [WR] classification can be assigned at this stage pending further follow-up. 

\subsubsection{Abell 48}
\label{A48}

Abell 48 is a highly reddened PN that was discovered by \citet{a3,a4}.  The central star exhibits a broad He II $\lambda$4686 line and a strong, broad 7118\AA\ feature.  Based on the presence of this feature, this could be mistaken for a [WC] in the AN03 scheme, but it is more likely a complex of six N IV lines between 7103\AA\ and 7129\AA\ if we take into account the recent WN identification of \citet{w4}.

The CS of Abell 48 was recently designated as a massive WN6 star in \citet{w4}, although our spectra predate this study.  However this object is more likely to be a PN than a massive WR ring nebula.  Abell 48 may well be another member of the rare [WN] or [WN/WC] class mentioned in \citet{t1}.  The arguments for this will be presented in a later paper (DePew et al. 2011, in preparation).  The spectrum, coordinates and observation details of this object can be found in Fig.~\ref{Spec3} and Tables \ref{ObsInfo} and \ref{New[WR]s}.  Abell 48 has been previously studied in the IR wavelengths by \citet{prl08}.  It is a clear detection in GLIMPSE \citep{cohen}.

\subsubsection{MPA1921+0132}

MPA1921+0132 exhibits a broad C IV $\lambda\lambda$5801,12 doublet (Fig.~\ref{Spec2}) typical of the [WR] class.  There is a broad, weaker feature centred around 4665\AA\ which may contain C IV $\lambda$4658 and He II $\lambda$4686. The continuum of this entire spectrum is undetected.  It is therefore impossible to determine if a line is truly absent.  We can only assert that C III $\lambda$5696 is weaker than C IV $\lambda\lambda$5801,12, making this an earlier spectral subtype than [WC8] in the CDB98 system.
 
\subsubsection{PHR0723+0036}

The relatively bright central star of this PN was originally identified as a WC star by \citet{d1}, who named it KPD 0721+0042.  \citet{d1} noted that the strongest two features in the spectrum of KPD 0721+0042 are the C IV lines.  He found no evidence of C III $\lambda$5696 emission, and consequently classified this object as WC4 following \citet{v1}.  \citet{d1} found the strength of the N III $\lambda$4634 line to be `unprecedented for any WC star'.  Furthermore, the estimated distance of 1200 pc above the Galactic plane argued for halo membership, leading \citeauthor{d1} to theorise that this was actually a Population II WC star, though he detected no nebular features from any associated nebula.  The surrounding PN was finally uncovered by the  MASH survey, indicating that the star is indeed a [WR].  The C IV $\lambda\lambda$5801,12 doublet in our spectrum (Fig.~ \ref{Spec2}) has an EW of $\sim$70\AA.  The feature around $\lambda$4650 is very broad, likely comprising C IV $\lambda$4658 and He II $\lambda$4686.  The continuum is evident.  Given the absence of C III $\lambda$5696 and oxygen emission lines, this star is likely a [WC4] in the CDB98 system.

Given the intensity of the C IV/He II blend at $\sim$4650\AA\ ($\sim$0.5 of F($\lambda\lambda$5801,12)) and the absence of C III $\lambda$5696, we confirm a [WC4] classification according to the AN03 scheme.

\subsubsection{PHR0654-1045}

PHR0654-1045 was first mentioned as a possible PN (PM 1-23) by \citet{p4} based on IRAS  photometry. It was similarly classified as a PN candidate by \citet{gl90} who designated it as GLMP 160. The object was unambiguously shown to be a PN in the MASH catalogue \citep{p1} based on spectroscopic confirmation in January 2003, where the [WR] CSPN features were first noted. The CSPN  was subsequently included in Acker \& Neiner (2003). \citet{pm03} independently noted [WR] features in the CSPN classifying it as a [WC6-7] star according to the CDB98 scheme.  It is also listed in the catalogue of \citet{s5} as a [WC7].  

Based on our data the central star exhibits the strong C IV doublet $\lambda\lambda$5801,12 with EW$\sim$7\AA, as well as a strong C III $\lambda$5696 line with EW$\sim$4\AA, implying an approximate classification of [WC7] according to the CDB98 scheme (see Figs. \ref{Spec1} to \ref{Spec3}).  This object displays a prominent He II line at 4686\AA.  O VI $\lambda\lambda$3811,34 are absent.  The strong broad feature at the wavelength of H$\alpha$ might be partly due to stellar He II.  The FWHM of the C IV doublet ($\sim$25\AA) is within range of the [WC7] spectral subtype in the CDB98 scheme.  The C III $\lambda$4650 line appears to be approximately 3.5 times as intense as the C IV $\lambda\lambda$5801,12 doublet.  The intensity of the C III $\lambda$5696 line ($\sim$0.45 of the C IV doublet) and the width of the C IV $\lambda\lambda$5801,12 doublet also argue for a [WC8] according to the AN03 scheme.

\citet{hzg} have recently found this star to be a photometric variable with a period of 0.63 days, possibly indicating a close binary.  This would be the only currently known [WR] star to have passed through a common envelope phase.  These authors also classified the CS as [WC7]. 

\subsubsection{Wray 17-23}

This PN was originally found by \citet{w3} and is noted in the catalogue of \citet{s5} as IRAS08574-5011.  Ionised oxygen species seem lacking from the spectra of the central star that we have obtained (see Figs. \ref{Spec1} to \ref{Spec3}).  The C IV $\lambda\lambda$5801,12 doublet is weak and broad (EW$\sim$25\AA, FWHM$\sim$50\AA), but C III $\lambda$5696 is not visible.  Since the continuum is barely detected, we estimate that the C III $\lambda$5696 line has to be weaker than $\sim$50~\% of the C IV doublet.  This suggests that this object has a spectral type of [WC7] or earlier according to CDB98 and [WC5-6] in the AN03 system.  [O~III] $\lambda\lambda$4959,5007 project from broad pedestals due to scattered light in the instrument.  
  
\subsubsection{PHR1001-6152}

The C IV $\lambda\lambda$5801,12 doublet is present.  The He II $\lambda$4686 line is very prominent in the central star's spectrum (Figs. \ref{Spec1} to \ref{Spec3}); He II $\lambda$5411 is not readily apparent.  O VI $\lambda\lambda$3811,34 is not apparent, but the continuum in this region is too noisy to conclude that the line is absent.  C III $\lambda$5696 is not evident.  The absence of C III $\lambda$5696 and the width of the C IV $\lambda\lambda$5801,12 doublet would argue for a [WC4] subtype or earlier in the CDB98 scheme.  However, the C IV $\lambda\lambda$5801,12 line (EW$\sim$10\AA) marks this object as having an unusually weak-lined spectrum.  We refrain from designating it a WELS because the C IV doublet is far broader than normal for this class \citep{m4}.

The He II $\lambda$4686 line is almost three times as intense as the C IV $\lambda\lambda$5801,12 doublet, but this is due more to the weakness of the doublet.  The possible O VI $\lambda$5290 line is $\sim$0.6 times the strength of the C IV doublet, but this measurement is very uncertain.  Based on the strengths of the He II $\lambda$4686 line and C IV $\lambda$4658 line, and the apparent absence of the high-excitation O VI $\lambda$3822 doublet, this star is classified as a [WO3] in the AN03 scheme.

\subsubsection{BMP1209-5553}

The central star of BMP1209-5553 shows a very strong O VI $\lambda\lambda$3811,34 doublet (Fig. \ref{Spec1}), which, along with the emission blend at 4650\AA, comprise the strongest stellar features seen in this spectrum.  The blend likely comprises C IV $\lambda$4658 and He II $\lambda$4686, the latter of which appears to possess a nebular component.  This star may exhibit a weak C IV $\lambda\lambda$5801,12 feature. The continuum is not observed for this star.  Based on the strength of the O VI $\lambda$3811 line, which is much stronger than both C IV $\lambda\lambda$5801,12 and O V $\lambda$5590, this star is likely a [WO1] in the CDB98 system, while based on the strength of He II $\lambda$4686 and O VI $\lambda$ 3811, and the weakness of C IV $\lambda\lambda$5801,12, this is a [WO2] in the AN03 system.  

\subsubsection{MPA1243-6428}

The C III $\lambda$5696 and C II $\lambda$7236 lines are prominent (Fig.~\ref{Spec3}), while C IV $\lambda\lambda$5801,12 are not observed.  C II $\lambda$4267 is in an extremely noisy part of the spectrum, so we cannot conclude that it is absent.  However, the continuum is not observed, so we cannot guarantee that C IV is not present.  Since C III $\lambda$5696 is clearly much stronger than C IV $\lambda\lambda$5801,12, we can deduce that this star has a later subtype than [WC9] in the CDB98 scheme and approximately [WC9] in the AN03 one.  Helium lines are absent.  The Balmer lines are present, but the nebular [O III] lines are not.  In [WC9-11]s, the ionisation of the PN would be low, so this is to be expected.  

\subsubsection{MPA1354-6337}

The broad C IV $\lambda\lambda$5801,12 doublet is strong (Fig. \ref{Spec2}).  The C III $\lambda$5696 line is not visible.  However, since the continuum is not detected, we can only say that this star has a spectral type earlier than [WC7] according to CDB98.  Because the C IV doublet is in any case more intense than C III $\lambda$5696, this star must be approximately a [WC4] in the AN03 system.

\subsubsection{PHR1416-5809}

This central star has strong C III $\lambda$5696, C IV $\lambda$4658 and He II $\lambda$4686.  The C III $\lambda$5696 line has an intensity over twice that of the C IV doublet, indicating that this object is a [WC9] in both CDB98 and AN03 schemes.

\subsubsection{MPA1611-4356}

\begin{figure*}
\begin{center}
\setlength{\intextsep}{1.0cm}
\begin{minipage}[h]{17cm}
\includegraphics[clip=true,scale=0.9]{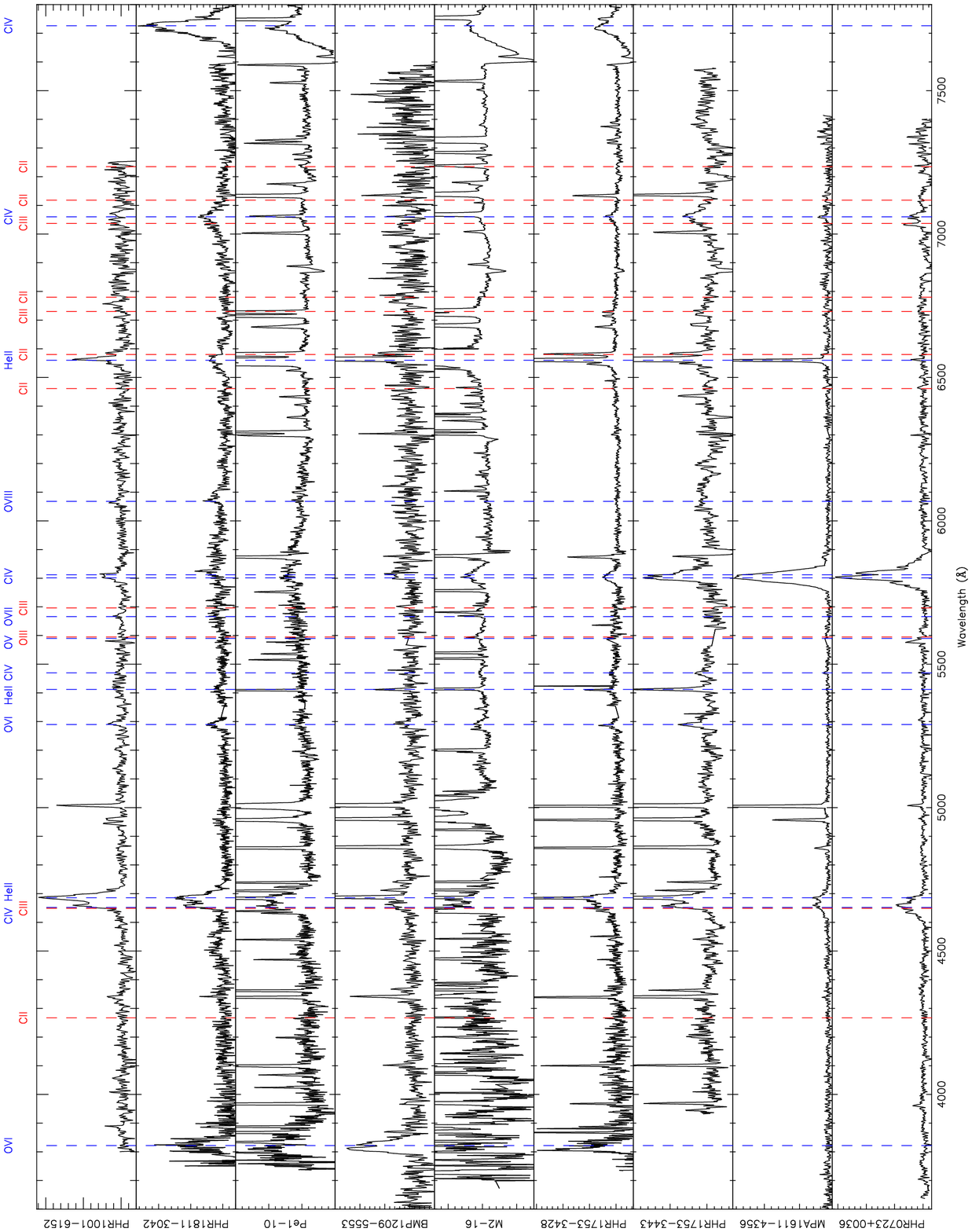}
\end{minipage}
\caption{\setlength{\baselineskip}{20pt}Spectra of objects whose central stars have recently been identified as being [WR]s or WELS;  all spectra have been rectified.  The most prominent lines have been identified (dashed lines and labels).}
\label{Spec1}
\end{center}
\end{figure*}

MPA1611-4356 first appeared in \citet{m1}.  The central star displays a strong, broad C IV $\lambda\lambda$5801,12 doublet, with a FWHM of $\sim$40\AA\ and an EW of $\sim$600\AA\ (Figs. \ref{Spec1} to \ref{Spec3}).  This is easily the strongest stellar feature in the spectrum.  He II $\lambda$4686 is heavily blended with the C IV $\lambda$4658 line.  This object has a low SNR but the strength of C IV $\lambda\lambda$5801,12, absence of C III $\lambda$5696, and presence of the C IV - He II blend at $\sim$4670\AA\ argues for a [WC4] or a [WO4] classification.  In a [WC/O4] the C IV - He II blend is stronger compared to C IV $\lambda\lambda$5801,12; however, the lack of a continuum detection in the blue makes the bluer of the two features artificially weak.  Yet it is strange that the He II $\lambda$4686 line appears stronger than C IV $\lambda$4658 in this star.  This is the opposite of what is expected in a normal [WC/O4] \citep[see][]{c1}.


\subsubsection{MPA1655-3535}

This object was classified as an early-type emission-line star by \citet{w1}, but the surrounding PN was not discovered until the MASH survey, likely owing to its compact nature.  C III $\lambda$5696 is detected (Fig.~\ref{Spec3}), as is a possible C II $\lambda$4267 line ($\sim$0.8 times the intensity of C III $\lambda$5696).  C IV $\lambda\lambda$5801,12 is absent.  Considering that the continuum signal-to-noise ratio is extremely low, we classify the star as later than a [WC10] in the CDB98 scheme (the C IV $\lambda\lambda$5801,12 doublet could be one-tenth the intensity of C III $\lambda$5696 and be lost in the continuum noise). There is no nebular [O III] $\lambda$5007 line, which, when combined with its small, compact nature and its Galactic latitude, indicate that this is a very low-excitation PN containing a cooler central star. 

Given the strong C II $\lambda$7236 line, this is probably later than a [WC10] in the AN03 system.  

\subsubsection{Pe 1-10}

Pe 1-10, also known as Hf 2-1, was first mentioned in \citet{p5}.
\citet{misz09b} presented an AA$\Omega$ spectrum of the central star which clearly showed broad, relatively strong emission features at O~VI $\lambda\lambda$3811,34 and around He II $\lambda$4686, indicating a [WO] classification.  This latter feature is likely blended with C IV $\lambda$4658.  Narrow nebular lines at He II $\lambda\lambda$4686,5411 and He I $\lambda$5876 are visible.  Broad weaker stellar C IV features may be present around 5806\AA\ and 7726\AA.  This star was also observed by \citet{g4} but their long-slit spectra either missed the CSPN or were too shallow to reveal the [WO] features.

\subsubsection{PHR1753-3443}
\label{PHR1753-3443}

This PN is now also known as PN K 6-32 \citep[see][]{k1}, but the MASH discovery and spectrum of the central star predates its publication in that catalogue.

While C IV $\lambda\lambda$5801,12 is present (EW$\sim$12\AA), C III $\lambda$5696 is not; this CS is probably hotter than [WC4] in the CDB98 classification.  C IV $\lambda$4658 is detected, while He II $\lambda$4686 has a nebular component.  The width of the C IV line is narrower than typical for a [WC4]. 

O VI $\lambda$5290 appears to be present ($\sim$0.57 of C IV $\lambda\lambda$5801,12), as is C IV $\lambda$7060 ($\sim$0.17 of C IV $\lambda\lambda$5801,12) suggesting that this is a [WO3] according to AN03.  

\subsubsection{PHR1753-3428}

The C IV $\lambda\lambda$5801,12 doublet is relatively strong.  The C IV - He II blend at 4650\AA\ is also present.  There is no C III $\lambda$5696, so this star is earlier than [WC4].  As is the case for PHR1753-3443, the FWHM of the C IV $\lambda\lambda$5801,12 is rather narrow for an early [WC] star.  

\subsubsection{Th 3-13}

Th 3-13 is listed as a WELS PN in \citet{g1}.  However, our spectrum presents a strong and broad C IV $\lambda\lambda$5801,12 feature.  The C III $\lambda$5696 feature is shallow but approximately as broad as the C IV doublet.  The intensity ratio of the C IV doublet to the C III $\lambda$5696 line ($\sim$3.9) falls within the range delineated for [WC6-7]s in the CDB98 scheme, although the width of the doublet is low for that class.  The C II $\lambda$4267 line is not visible, but this may be due to reddening, which is large in this object (c(H$\beta$)=3.6).  

The C III $\lambda$5696 line is 0.27 times as intense as the C IV doublet and the C III-IV $\lambda$4650 blend is almost three times the value of the doublet, suggesting a classification of [WC5-6] in the AN03 scheme.

\subsubsection{PHR1752-3330}

The C IV $\lambda\lambda$5801,12 doublet is absent, as is C II $\lambda$4267.  C III $\lambda$5696 is present, as is C II $\lambda$7236, but both are very weak and narrow.  A system of absorption lines appears at $\sim$5800\AA, as well as $\sim$4650\AA.  In the CDB98 scheme, this object cannot be classified as a late [WC] because the C III line at 5696\AA\ is too weak {\it and} both C IV $\lambda\lambda$5801,12 and C II $\lambda$4267 are lacking.  CDB98 would call this an emission-line star.  Due to the absorption line at He II $\lambda$5411, this is a [WC11] according to the AN03 scheme.  

\subsubsection{M 2-16}

This PN is listed as a non-emission-line star in \citet{g3}.  However, the central star is classified as a variable Be candidate by \citet{s2}, though no further explanatory details are provided.  A broad C IV $\lambda\lambda$5801,12 doublet and C IV $\lambda$7726, as well as a possible O V $\lambda$5590 are detected.  This is hotter than [WC4] in the CDB98 scheme.

\subsubsection{PHR1731-2709}

\begin{figure*}
\begin{center}
\setlength{\intextsep}{1.0cm}
\begin{minipage}[h]{17cm}
\includegraphics[clip=true,scale=0.9]{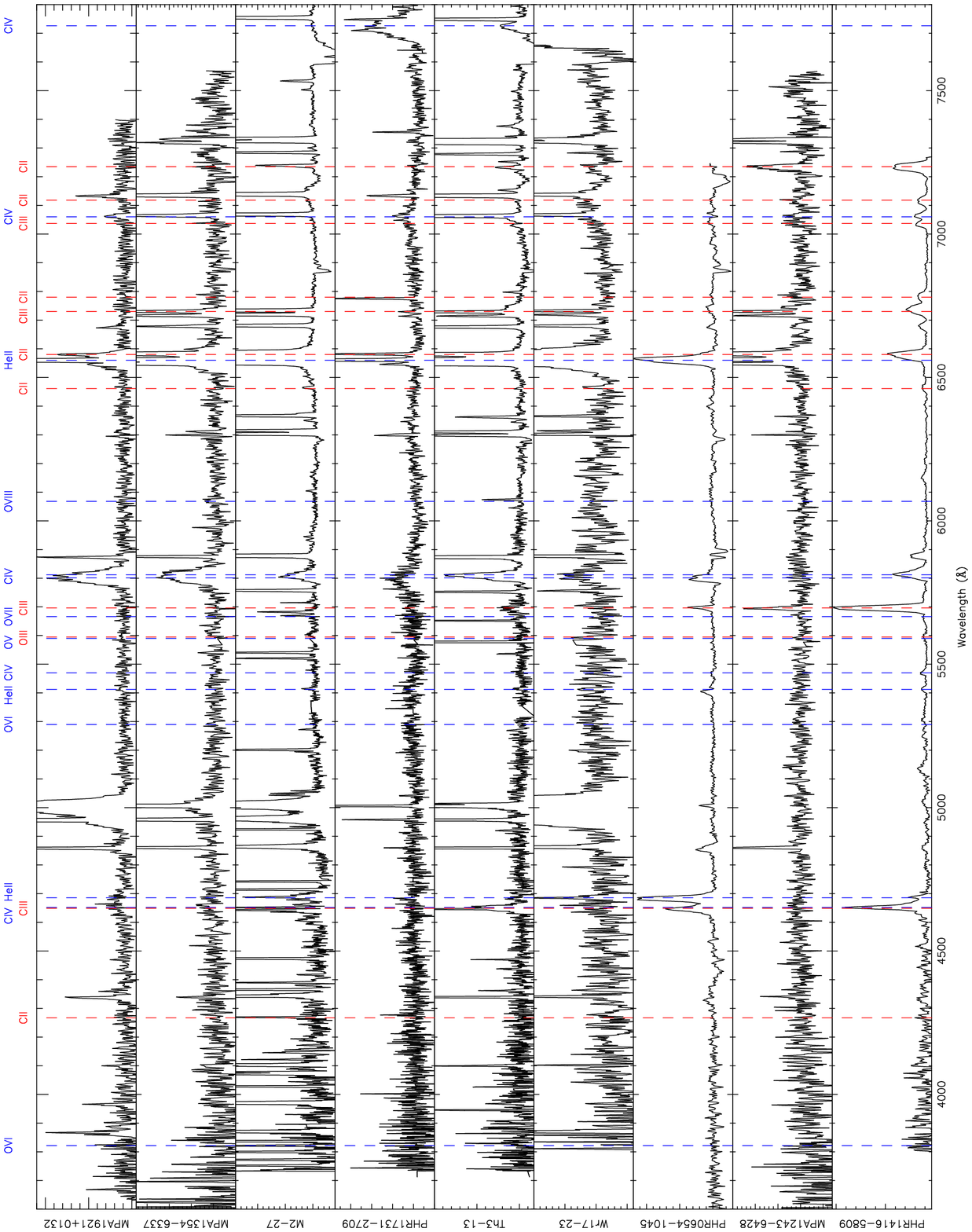}
\end{minipage}
\caption{\setlength{\baselineskip}{20pt}Spectra of objects whose central stars have recently been identified as being [WR]s or WELS; all spectra have been rectified.  The most prominent lines have been identified (dashed lines and labels).}
\label{Spec2}
\end{center}
\end{figure*}

The [WR] character of the reddened central star of PHR1731-2709 was originally overlooked in the PN confirmatory spectrum.  More recent observations in May 2008 have uncovered a broad feature around C IV $\lambda\lambda$5801,12 (Fig. \ref{Spec2}).  C IV $\lambda$7726 is also detected (not shown in the figure).  The lack of continuum detection allows us to say only that C IV $\lambda\lambda$5801,12 is stronger than C III $\lambda$5696.  Therefore, this star has an earlier type than [WC8] in the CDB98 system.  

\subsubsection{M 2-27}

The CS of M 2-27 was originally classified as a WELS in \citet{g1}.  However, this object has a weak but broad (EW$\sim$15\AA, FWHM$\sim$25\AA) C IV $\lambda\lambda$5801,12 doublet in our spectra.  Neither oxygen excitation lines nor the C III $\lambda$5696 line appear to be present, although C II $\lambda$7236 is visible.  Due to the weakness of C III $\lambda$5696, this is likely earlier than a [WC8] in the CDB98 scheme.  

\subsection{Possible [WR]s and WELS}

In this section we briefly detail the spectra of several objects that may exhibit [WR] features, including WELS, but require deeper spectra to allow a proper classification.

\subsubsection{M 2-21}

M 2-21 was first mentioned in \citet{w3}, where it was designated Wr 16-337.  The central star was designated a WELS in \citet{g4}, but our observations predate this publication.  This object displays partially resolved C IV doublet $\lambda\lambda$5801,12 lines and a N III/C III-IV complex at $\lambda\lambda$4631,4641,4650.  C III $\lambda$5696 is absent, in line with a WELS classification.  The presence of C II lines is surprising, given the absence of C III and the presence of C IV, suggesting a different origin, possibly nebular.    

\subsubsection{M 1-35}

M 1-35 is a Type I bulge PN.  Interestingly, it was observed by \citet{gs03}, but no emission lines were found; however, we detect an emission-line spectrum.  The N III/C III-IV complex around $\lambda$4640 is prominent.  The lines appear to be narrow.  The C IV $\lambda$5801,12 doublet is a doubtful detection, but if present is very faint.  C III $\lambda$5696 is absent, so this is potentially a WELS.  However, the weakness/absence of C IV $\lambda\lambda$5801,12, combined with the strength of the N III/C III-IV complex, potentially indicates an alternative identification as an irradiated close binary central star.

\subsubsection{M 2-39}

M 2-39 is classified as a WELS in \citet{g1} and is mentioned in \citet{g4} as a low metallicity PN.  Our MASH observation of this object predates its publication in \citet{g1}.

The broad C IV doublet has a FWHM of 20\AA\ and an EW of 5\AA\ (Fig. \ref{Spec3}).  C III $\lambda$5696 is not visible.  He II $\lambda$4686 is present, but He II $\lambda$5411 is not seen.  N III $\lambda$4634 and $\lambda$4641 are present alongside C III $\lambda$4650 and C~IV $\lambda$4658.

Based on the high ionisation of this star, as well as the narrowness of the observed emission lines, this star is probably a WELS.

\subsubsection{M 2-42}

M 2-42's C IV $\lambda$5801,12 doublet lines are easily visible with an EW=13\AA.  However, C II $\lambda$4267 and C III $\lambda$5696 are not observed.  The He II lines at 4686 and 5411\AA\ are relatively strong.  We observe the 4650-\AA\ line complex, although the identification of the specific lines is not certain.  We fit two lines at $\sim$4607 and $\sim$4641\AA, which could correspond to the N V $\lambda\lambda$4604,20 doublet and N III $\lambda$4634,41 doublet.  Beyond these, there are no visible stellar lines.  This object could be a WELS.

\subsubsection{MPA1405-5507}

\begin{figure*}
\setlength{\intextsep}{1.0cm}
\begin{minipage}[t]{17cm}
\begin{center}
\includegraphics[clip=true,scale=0.9]{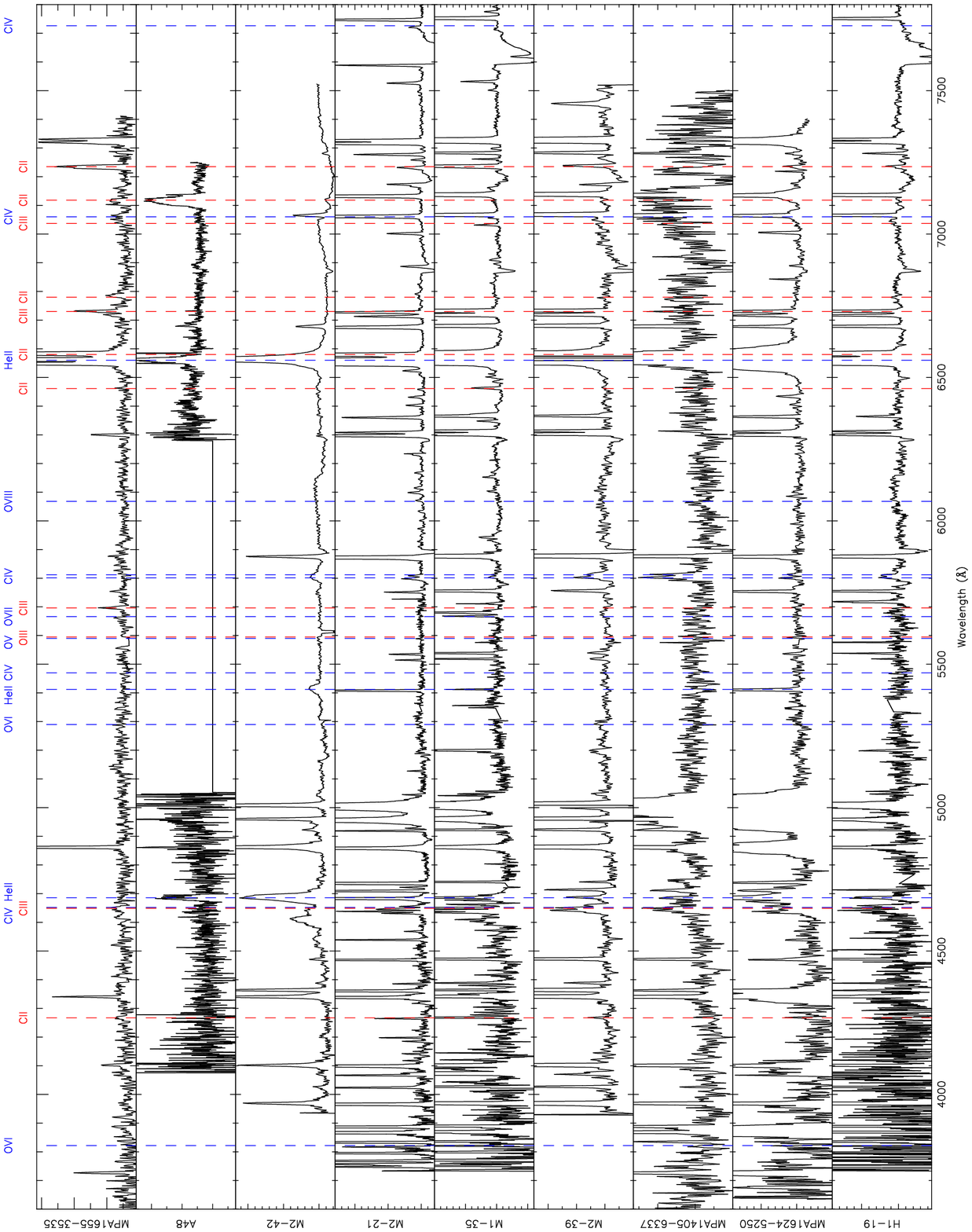} 
\end{center}
\end{minipage}
\caption{\setlength{\baselineskip}{20pt}Spectra of objects whose central stars have recently been identified as being [WR]s or WELS; all spectra have been rectified.  The most prominent lines have been identified (dashed lines and labels).}
\begin{flushleft}
\end{flushleft}
\label{Spec3}
\end{figure*}

The central star is classed as [WC4-6] in the MASH-II catalogue.  C IV $\lambda\lambda$5801,12 is present and He II $\lambda$4686 is weak (Fig. \ref{Spec3}), while C III $\lambda$5696 is not visible.  No oxygen lines are detected.  The 4650\AA\ emission complex is present.  

The ratio of C IV $\lambda\lambda$5801,12 to C III $\lambda$5696 is likely larger than 10, making this an earlier spectral subtype than [WC5] in the CDB98 system.
However, given the narrowness of the C IV $\lambda\lambda$5801,12 and the lines in the $\lambda$4650 complex, this object could also be a WELS.

\subsubsection{Pe 2-8}

This PN, originally noted in \citet{p5}, has been previously observed by \citet{dh97} and \citet{h2}.  In \citet{h2}, Pe 2-8 was observed in IR with Spitzer and used for comparison against several LMC PNe.

The stellar continuum is barely detected.  The C IV doublet exhibits separate peaks at 5801 and 5812\AA.  This raises the possibility of a WELS identification, although it must be noted that the EW of both peaks (12 and 12\AA\ respectively) are larger than normal for WELS.  We cannot speculate further since our spectral range does not contain the other diagnostic lines (e.g. C III $\lambda$5696).

\subsubsection{MPA1624-5250}

The C IV $\lambda$5801 and $\lambda$5812 lines each have an equivalent width approximately equal to the instrumental resolution of 5\AA.  C III $\lambda$5696 is absent.  There is a faint but visible N III/C III-IV complex around 4650\AA.  This object is likely a WELS.

\subsubsection{M 3-38}

This object exhibits both C IV $\lambda$5801 and $\lambda$5812, but these lines are narrow and well separated.  The C III $\lambda$5696 line was not in the observed spectral range, but the prominent N III $\lambda$4634 and $\lambda$4641 lines indicate that this is a WELS.

\subsubsection{Sa 3-107} 

The individual peaks of the C IV $\lambda\lambda$5801,12 doublet are visible, a characteristic of many late [WC]s and WELS.  This doublet has an EW$\sim$9\AA, which is also a common characteristic of WELS.  The C IV $\lambda$7726 line is also present.  O VI $\lambda\lambda$3811,34 and C II $\lambda$4267 are absent.  C III $\lambda$5696 is absent, which suggests that this star is not a late-type [WC], but a WELS.





\subsubsection{H 1-19}

H 1-19 is a bulge object.  Its chemical abundances have been previously examined in \citet{cgsb}.  The central star exhibits a partially resolved C IV $\lambda\lambda$5801,12 doublet and the N III/C III-IV complex at $\sim$4640\AA.  In addition, C III $\lambda$5696 is absent, indicating this to be a WELS. 

\begin{table*}
\caption{A list of the newly discovered [WR] CSPNe, along with WELS found in the course of examining the sample.  The top portion lists all of those PNe from the MASH sample, while those objects below the line are for previously known PNe.}
\begin{tabular}{l l l l c c c c c c c c c}
\hline
PN G & Name & Subclass$^{1}$ & Subclass$^{1}$ & Morph.$^{2}$ & Sample$^{3}$ & c(H$\beta$) & V & I & J & K$_{S}$ & Phot.\\
 & & (CDB98) & (AN03) & & & & & & & & Source$^{4}$ \\ [0.5ex]
\hline
\hline
001.2-05.6 & PHR1811-3042 & [WO] & [WO1-2] & E & M-I & 1.37 & -- & -- & -- & 12.28 & U \\
016.8-01.7 & BMP1827-1504 & [WR]: & [WR] & R & M-II & 3.69 & -- & 15.41 & 14.46 & 13.03 & D \\
037.7-06.0 & MPA1921+0132 & $\leq$[WC8] & [WC4]: & R & M-II & 1.17 & 16.6 & 17.30 & $>$15.83 & 14.89 & N \\
216.0+07.4 & PHR0723+0036 & [WC4]: & [WC4]: & E & M-I & 0.32 & 13.9 & 13.97 & 13.90 & 13.83 & D84,F08 \\
222.8-04.2 & PHR0654-1045 & [WC7] & [WC8] & E & M-I & 1.13 & 14.1 & 13.02 & 12.28 & 11.65 & N,D \\
284.2-05.3 & PHR1001-6152 & $\leq$[WC4] & [WO3] & R & M-I & 1.11 & 16.4 & -- & 16.21 & -- & N,G \\
297.0+06.5 & BMP1209-5553 & [WO1]: & [WO2]: & B & M-II & 0.38 & -- & -- & -- & -- & -- \\
302.0-01.6 & MPA1243-6428 & $\geq$[WC9] & [WC9]: & S & M-II & 2.71 & -- & 16.74$^{*}$ & 14.84 & 12.61 & D \\
309.8-01.6 & MPA1354-6337 & $\leq$[WC7] & [WC4]: & E & M-II & 2.85 & -- & 17.00 & 14.97 & 13.50 & G \\
313.4+06.2 & MPA1405-5507 & $\leq$[WC5]/WELS & WELS & S & M-II & 0.37 & 15.8 & -- & 15.57 & 14.76 & N,G \\
313.9+02.8 & PHR1416-5809 & [WC9] & [WC9] & E & M-I & 1.79 & 16.2 & 14.97$^{*}$ & 13.74: & 12.74 & N \\
331.8-02.3 & MPA1624-5250 & WELS & WELS & R & M-II & 1.88 & 16.5 & 15.96$^{*}$ & 14.26 & 12.16 & N,D \\
336.5+05.5 & MPA1611-4356 & [WC/O4]: & [WO4] & R & M-II & 0.80 & 17.4 & 17.12$^{*}$ & 16.36 & 15.64 & N,G \\
348.4+04.9 & MPA1655-3535 & $\geq$[WC10] & [WC10] & S & M-II & 1.34 & 16.4$^{5}$ & 15.65 & 14.66 & 13.17 & N,G \\
355.9-04.4 & PHR1753-3443$^{6}$ & $\leq$[WC4] & [WO3]: & B & M-I & 2.36 & 15.7 & -- & 13.22 & 12.09 & G \\
356.0-04.2 & PHR1753-3428 & $\leq$[WC4] & [WO2-3] & E & M-I & 2.20 & 17.8 & -- & -- & -- & G \\
356.8-03.6 & PHR1752-3330 & EL & [WC11] & E & MU & 1.32 & 14.4 & 14.32 & 13.28 & 12.83 & N,D \\
359.8+03.5 & PHR1731-2709 & $\leq$[WC8] & [WC4]: & E & M-I & 3.31 & -- & -- & -- & -- & -- \\
\hline
000.7-02.7 & M 2-21 & WELS & WELS & R & SEC & 0.60 & 13.2 & -- & 13.20 & 12.02 & N \\
003.9-02.3 & M 1-35 & WELS: & WELS & R & SEC & 1.36 & 15.1 & 14.66 & 12.95 & 11.87 & N,D \\
008.1-04.7 & M 2-39 & WELS & WELS & R & SEC & 1.67 & 13.7$^{\dagger}$ & 14.51 & 13.53 & 12.04 & N,D \\
008.2-04.8 & M 2-42 & WELS: & WO:: & E & SEC & 2.40 & 15.2 & 15.29 & 14.45 & 13.53 & N,D \\
029.0+00.4 & Abell 48$^{7}$ & -- & WC:: & E & SEC & 3.02 & 17.7 & 15.50 & 13.51 & 12.33 & N,D \\
270.1-02.9 & Wr 17-23 & $\leq$[WC7] & [WC5-6] & E & SEC & 2.51 & 16.7 & 16.48$^{*}$ & 13.97 & 12.27 & G \\
322.4-00.1 & Pe 2-8 & [WR]/WELS & WELS & E & SEC & 2.96 & 17.3 & 14.80$^{*}$ & 11.81 & 9.96 & G \\
355.4-04.0 & Pe 1-10 & [WO]:: & [WO2] & E & SEC & 1.63 & -- & -- & -- & -- & -- \\
356.1+02.7 & Th 3-13 & [WC6-7] & [WC5-6] & R & SEC & 3.59 & -- & 16.14 & 13.67 & 11.95 & D \\
356.9+04.4 & M 3-38 & WELS & WELS & R & SEC & 1.78 & 17.2 & 15.74$^{*}$ & 13.97 & 12.52 & N,D \\
357.4-03.2 & M 2-16 & $\leq$[WC4] & [WO2-3] & E & SEC & 0.80 & 16.5 & 15.18 & 13.68 & 12.56 & S08 \\
358.0-04.6 & Sa 3-107 & WELS & WELS & R & SEC & 1.43 & 16.4 & 15.80 & -- & -- & G \\
358.9+03.3 & H 1-19 & WELS & WELS & R & SEC & 2.04 & 17.1 & 14.63 & 12.39 & 11.05 & N,D \\
359.9-04.5 & M 2-27 & $\leq$[WC8]: & [WC4]: & R & SEC & 0.94 & 14.4$^{\dagger}$ & 14.32 & 12.93 & 11.78 & D \\
\hline
\hline
\end{tabular}
\label{New[WR]s}
\begin{flushleft}
$^{1}$[WR] subclasses are very approximate because of the modest quality of our spectra.  Where ``WELS'' is indicated, this object fails the respective [WR] subclass scheme and meets the criteria listed in \S\ref{ID} for that class. ``EL'' indicates that emission lines at WR wavelengths were seen, but the object does not meet the other criteria for [WR] classification.\\
$^{2}$B=Bipolar, E=Elliptical, I=Irregular, R=Round, S=quasi-Stellar.\\
$^{3}$M-I=MASH-I \citep{p1}, M-II=MASH-II \citep{m1}, MU=Found in the MASH sample but previously unpublished, SEC=PN previously known from Strasbourg-ESO Catalogue \citep{a1}.\\  
$^{4}$J and K$_{S}$ magnitudes are from 2MASS\citep{2mass} unless otherwise indicated.  N=NOMAD-I \citep{z1}, G=GSC 2.3.2 \citep{l1}, D=DENIS \citep{d3}, D84=\citet{d1}, F08=\citet{f2}, S08=\citet{s2}, U=UKIDSS \citep{ukidss}.\\
$^{5}$V magnitude given as 13.1 in \citet{w1}.  This measurement is probably contaminated by a nebular component.\\
$^{6}$Also called K 6-32 in \citet{k1}.  See \S \ref{PHR1753-3443}.\\
$^{7}$Probable [WN/WC].  See \S\ref{A48} for a discussion of the classification.\\
$:$Classification uncertain. \\
$^{*}$Mean of two observations.\\
$^{\dagger}$Nebular contamination.
\end{flushleft}
\end{table*}

\section{Conclusions}

Thirty-two new [WR] CSPNe, candidate [WR]s and WELS have been presented.  Twenty-two members in this sample are identified as probable [WR]s; the remaining 10 appear to be WELS.  While 18 of these objects have been discovered in the MASH sample, 14 were found in previously known PNe through serendipitous fibre placement during follow-up observations of MASH objects with wide-field MOS systems such as AA$\Omega$, 2dF and 6dF.  This illustrates first and foremost that the central stars of more planetary nebulae should be observed.  Estimates of the percentage of CSPNe known to exhibit WR features have in the past hovered around 7--9 per cent \citep{tyl,t96,a2,frew10b}.  Hydrogen-deficient stellar evolution is still poorly understood, but its understanding may become more urgent if the proportion of stars which display this phenomenon grows.  A survey of a large sample of CSPNe may help to place theoretical constraints on stellar evolution models.  


Proper subclass designations will require deeper observations of the 32 CSPNe mentioned here.  The spectra presented in this paper are preliminary and necessitate follow-up.  A future paper will analyse the Galactic distribution and dynamical age of these newly discovered objects (DePew et al. 2011, in preparation).

\section*{Acknowledgments}

KD and AVK acknowledge receipt of MQRES PhD scholarships from Macquarie University.  The authors thank Donna Burton for service observations on the UK Schmidt Telescope and Stephen Fine and Scott Croom for observations of Sa~3-107 and M 2-16 with AA$\Omega$.  The authors also thank the anonymous referee for many valuable comments and suggestions. 

We made use of the Vizier service and SIMBAD database from the Centre de Donn\'ees astronomiques de Strasbourg, France.  This publication makes use of data products from the Two Micron All Sky Survey, which is a joint project of the University of Massachusetts and the Infrared Processing and Analysis Center/California Institute of Technology, funded by the National Aeronautics and Space Administration and the National Science Foundation.

\section*{Appendix A}

Here we mention five objects that possess interesting broad-lined spectra which we have discovered in the course of our analysis of the MASH database.  These objects were originally selected on the preliminary basis of having exhibited lines at or near [WR] wavelengths.  On deeper examination, they appear to lie outside the typical [WC]/[WO]/WELS classifications.  They are detailed below.

\subsection*{PHR1757-1649}

This is an unusual, high-excitation nebula with a quite irregular morphology.  Our SHS images show a smaller limb-brightened, coherent elliptical shell in the NE portion, surrounding a prominent central star.  We interpret this as a possible evolved PN embedded in a larger unrelated H{\sc II} region, in turn ionised by UV photons leaking out of the optically-thin PN shell.  Our spectra of the central star show WN features. However, for any reasonable reddening, this object would be very remote ($D>$ 10 kpc) if the nucleus was a Population I WR star.  The lack of warm dust seen in MSX images combined with the large $z$ height at the assumed distance militates against an interpretation as a star-forming region.  We hence conclude that this may be a PN around a H-deficient low-mass star, possibly a new member of the [WN] or [WN/WC] class.  There appear to be no carbon lines, however, so we refrain from designating it as a probable [WR] star.  Further observations are planned for this interesting object.

\subsection*{MPA1518-4738}

The only features in the spectrum of this object are the broad He II $\lambda$4686, H$\alpha$ and H$\beta$ lines.  There is an emission blend at 4650\AA, which could be the N III, C III-IV complex.   The continuum is not detected.  The FWHM of H$\alpha$, H$\beta$, and He II $\lambda$4686 are $\sim$1400 km s$^{-1}$ broader than the spectral resolution and much broader than would be expected for nebular lines, indicating a stellar origin.  

\subsection*{MPA1602-5543}

Our spectrum of this object is very poor; the one prominent line is a broad feature centered on 6558\AA.  This may be a blend of [N II] $\lambda\lambda$6548,6584, He II $\lambda$6560 and/or H$\alpha$.  We refrain from classifying this star until better spectra are obtained.

\subsection*{PHR1603-5402}
The emission line star is part of a resolved triple star on SuperCOSMOS Sky Survey\footnote{http://www-wfau.roe.ac.uk/sss/} images.  The surrounding nebula was classified as a reflection nebula (GN 15.59.8) by \citet{mag}, but it appears to be a compact H{\sc II} region holding a massive WR star.  This would have to be a subclass earlier than WC8 considering the wide (FWHM$\sim$130\AA) C IV $\lambda\lambda$5801,12 doublet and absence of C III $\lambda$5696.  The width of the C IV doublet would be typical of a WO2 star.  It deserves follow-up observations.

\subsection*{PPA1702-3509}

This appears to be a compact emission-line nebula with a 16th magnitude CS.  There is enough of a discrepancy between the published visual magnitudes from GSC \citep{l1} and NOMAD \citep{z1} and the DENIS \citep{d3} and 2MASS \citep{2mass} NIR photometry, to suggest that the CS may be variable.  Indeed the spectral features are reminiscent of V~Sge, the prototype of a very rare class of nova-like cataclysmic variables \citep{sd98} showing broad-line spectra.  The V~Sge stars show broad emission lines due to He II, O VI and N V.  Our spectra do not go far enough to the blue to detect the O VI $\lambda\lambda$3811,34 blend, but we detect a weak N V $\lambda$4603 line as well as a very broad $\lambda$4650 feature likely due to a blend of N III and C III-IV lines, and broad, bright He II $\lambda$4686, along with the Balmer lines of hydrogen.  There is a narrow [O~III] $\lambda$5007 line which shows there is a surrounding nebula present.  Its location makes it a possible outlying member of the Galactic bulge, and it warrants further study.


\begin{table*}
\caption{Observation details of the objects listed in Appendix A.}
\begin{tabular}{l c c c c c c c c c c}
\hline
PN G & RA & Dec & Name & Telescope & Instrument & Exposure & Grating & R & SNR$^{1}$ & Date \\
 & & & &   & & & & & & Observed \\ [0.5ex]
\hline
\hline
011.8+03.7 & 17:57:39.6 & -16:49:19 & PHR1757-1649 & SSO/2.3m & DBS & 1200 & 600B/600R & 1800 & 15 & 24 May 2006 \\
326.9+08.2 & 15:18:18.2 & -47:38:28 & MPA1518-4738 & SSO/2.3m & DBS & 300 & 300B & 1200 & -- & 24 Feb 2007 \\
327.5-02.2 & 16:02:11.2 & -55:43:30 & MPA1602-5543 & SSO/2.3m & DBS & 600 & 1200R & 600 & 5 & 06 May 2007 \\
328.8-01.1 & 16:03:41.4 & -54:02:04 & PHR1603-5402 & ESO/1.5m & B\&C & 1200 & 25 & 800 & -- & 03 Jul 2000 \\
349.7+04.0 & 17:02:46.1 & -35:09:02 & PPA1702-3509 & SSO/UKST & 6dF & 1200 & 580V & 1000 & 20 & 16 Aug 2004\\
\hline
\hline
\end{tabular}
\label{Not[WR]s}
\begin{flushleft}
$^{1}$SNR was measured in the continuum at $\sim$5900\AA.  When no SNR is indicated, this means that the continuum was not detected.\\
\end{flushleft}
\end{table*}




\label{lastpage}
\section*{Appendix B}

Table \ref{LineRat} presents FWHM, EW and relative intensities of our discovery [WR] and WELS spectra.

\begin{landscape}
\begin{table}
\caption{Below are FWHM, EW and dereddened intensities of stellar lines in our discovery spectra.  The FWHM and EW of C IV $\lambda\lambda$5801,12 and C III $\lambda$5696 are marked.  All other columns are the intensities of the lines, with C IV $\lambda\lambda$5801,12 = 100.  (We do not list the absolute C IV line fluxes because our spectra were not absolutely flux calibrated.)  Spaces marked `--' were not seen in the spectra.  `N.O.' indicates that the designated line lay in a region of the spectrum which was not observed.  `P' indicates that the line is present, but the exact value cannot be measured due to the absence of the continuum or C IV $\lambda\lambda$ 5801,12.  `S' indicates a strong line.  `W' signifies a weak line.  Again, `:' denotes an uncertain value, while `::' indicates a very uncertain value.}
\begin{tabular}{l c c c c c c c c c c c c}
\hline
Object Name & C IV $\lambda\lambda$5801,12 & C IV $\lambda\lambda$5801,12 & C III $\lambda$5696 & C III $\lambda$5696 & C III & O VI & C II & He II & O VI & O V & O VII & C II \\
 & FWHM (\AA)$^{*}$ & EW (\AA) & FWHM (\AA) & EW (\AA) & $\lambda$5696 & $\lambda\lambda$3811,34 & $\lambda$4267 & $\lambda$4686 & $\lambda$5290 & $\lambda$5590 & $\lambda$5670 & $\lambda$7236 \\ [0.5ex]
\hline
\hline
PHR1811-3042 & -- & -- & -- & -- & -- & -- & -- & P & P & -- & -- & -- \\
BMP1827-1504 & 3 & P & -- & -- & N.O. & N.O. & -- & -- & N.O. & N.O. & N.O. & -- \\
MPA1921+0132 & 42 & P & -- & -- & -- & -- & -- & -- & -- & -- & -- & -- \\
PHR0723+0036 & 47 & 65 & -- & -- & -- & -- & -- & 68$^{1}$ & -- & -- & -- & -- \\
PHR0654-1045 & 25 & 7 & 9 & 4 & 46 & -- & -- & 450$^{1}$ & -- & -- & -- & -- \\
PHR1001-6152 & 27 & 10 & -- & -- & -- & -- & -- & 1100:: & 54:: & -- & 13:: & -- \\
BMP1209-5553 & -- & -- & -- & -- & -- & S & -- & S & P & -- & -- & -- \\
MPA1243-6428 & -- & -- & 9 & P & P & -- & -- & -- & -- & -- & -- & P \\
MPA1354-6337 & 45 & S & -- & -- & -- & -- & -- & -- & -- & -- & -- & -- \\
MPA1405-5507 & 22: & 6: & -- & -- & -- & -- & -- & 37:: & -- & -- & -- & -- \\
PHR1416-5809 & 27 & 98 & 19 & 240 & 245 & -- & -- & 63$^{1}$ & -- & W & -- & 67: \\
MPA1624-5250 & 6,8: & P & -- & -- & -- & -- & -- & -- & -- & -- & -- & -- \\
MPA1611-4356 & 40 & 600 & -- &-- & -- & -- & -- & 17$^{1}$ & -- & -- & -- & -- \\
MPA1655-3535 & -- & -- & 7 & 11 & P & -- & -- & -- & -- & -- & -- & P \\
PHR1753-3443 & 29 & 12 & -- & -- & -- & N.O. & -- & 510$^{1}$ & W: & -- & -- & -- \\
PHR1753-3428 & 38 & 16 & -- & -- & -- & N.O. & W: & 330:$^{1}$ & -- & -- & -- & -- \\
PHR1752-3330 & -- & -- & 6 & 1 & P & -- & -- & -- & -- & -- & -- & P \\
PHR1731-2709 & 60 & P & -- & -- & -- & -- & -- & -- & -- & -- & -- & -- \\
\hline
M 2-21 & 7,5 & P & -- & -- & -- & -- & P & S & -- & -- & -- & P$^{1}$ \\
M 1-35 & -- & W: & -- & -- & -- & -- & P & P & -- & -- & -- & P \\
M 2-39 & 20 & 5 & -- & -- & -- & N.O. & W: & 16 & 14:: & -- & -- & 120 \\
M 2-42 & 25 & 5 & -- & -- & -- & N.O. & -- & 900 & -- & -- & -- & -- \\
A 48 & N.O. & N.O. & N.O. & N.O. & N.O. & N.O. & -- & P & N.O. & N.O. & N.O. & -- \\
Wr 17-23 & 50:: & 25:: & -- & -- & -- & -- & -- & W: & -- & -- & -- & -- \\
Pe 2-8 & 12,12 & 50:: & N.O. & N.O. & N.O. & N.O. & N.O. & -- & N.O. & N.O. & N.O. & P$^{1}$ \\
Pe 1-10 & 53 & 5 & -- & -- & -- & -- & -- & 4100$^{2}$ & W: & 17:: & -- & -- \\
Th 3-13 & 23 & P & 30 & P & P & -- & -- & -- & -- & -- & -- & P \\
M 3-38 & 4,2 & P & N.O. & N.O. & N.O. & N.O. & -- & S & N.O. & N.O. & N.O. & P \\
M 2-16 & 59 & 17 & -- & -- & -- & N.O. & N.O. & N.O. & W: & -- & -- & -- \\
Sa 3-107 & 7,7 & 9 & -- & -- & -- & N.O. & N.O. & N.O. & -- & -- & -- & 12 \\
H 1-19 & 17 & 5 & -- & -- & -- & -- & -- & 200: & -- & -- & -- & 35: \\
M 2-27 & 26 & 15 & -- & -- & -- & -- & 37: & 48: & -- & -- & -- & 66: \\
\hline
\hline
\end{tabular}
\label{LineRat}
\begin{flushleft}
$^{1}$Blended with other stellar and nebular lines.\\
$^{2}$Significant nebular component.\\
$^{*}$When two numbers are listed, individual components of the doublet were measured.
\end{flushleft}
\end{table}
\end{landscape}

\end{document}